\documentclass[12pt]{article}
\begin{document}
\pagestyle{plain}
\setcounter{page}{1}

\date{\today }

\begin{center}
{\large \textbf{Satellite E\"{o}tv\"{o}s Test of the Weak Equivalence
Principle for Zero-point Vacuum Energy}} \vskip 0.2 true in {\large J. W.
Moffat}\footnote{%
e-mail:john.moffat@utoronto.ca} \vskip 0.2 true in \textit{Department of
Physics, University of Toronto, Toronto, Ontario M5S 1A7, Canada} \vskip0.3
true in and \vskip0.2 true in \textit{Perimeter Institute for Theoretical
Physics, Waterloo, Ontario N2J 2W9, Canada} \vskip 0.2 true in and \vskip %
0.2 true in {\large G. T. Gillies}\footnote{%
e-mail: gtg@virginia.edu}, \vskip 0.2 true in \textit{School of Engineering
and Applied Science, University of Virginia, Charlottesville, VA 22904-4746,
USA}
\begin{abstract}
An E\"{o}tv\"{o}s experiment to test the weak equivalence principle (WEP)
for zero-point vacuum energy is proposed using a satellite. Following the
suggestion of Ross for a terrestrial experiment of this type, the
acceleration of a spherical test mass of aluminum would be compared with
that of a similar test mass made from another material. The estimated ratio
of the zero-point vacuum energy density inside the aluminum sphere to the
rest mass energy density is $\sim 1.6\times 10^{-14}$, which would allow a
1\% resolution of a potential WEP violation observed in a satellite mission
test that had a baseline sensitivity to WEP violations of $\sim 10^{-16}$.
An observed violation of the WEP for vacuum energy density would constitute
a significant clue as to the origin of the cosmological constant and the
source of dark energy, and test a recently proposed resolution of the
cosmological constant problem, based on a model of nonlocal quantum gravity
and quantum field theory.
\end{abstract}
\vskip 0.3 true in
\end{center}

\vskip 0.3 true in 

A recent proposal to resolve the cosmological constant problem was based on
a nonlocal quantum gravity theory and quantum field theory~\cite%
{Moffat,Moffat2}. The resolution demands that there exist a fundamental
low-energy gravitational energy scale, $\Lambda_{\mathrm{Gvac}}\sim 10^{-3}$
eV, above which the coupling of gravitons to vacuum energy density was
suppressed by a nonlocal gravitational form factor in the calculation of
standard model vacuum polarization loop graphs. This would require a strong
violation of the weak equivalence principle (WEP) for coupling of gravitons
to pure vacuum energy loop graphs compared to ordinary matter.

It is generally believed that a solution to the cosmological constant
problem could alter in a significant way our understanding of gravitation
and particle physics~\cite{Straumann}. There is mounting observational
evidence~\cite{Perlmutter} that the universe is accelerating and that a form
of dark energy exists. A small cosmological constant corresponding to a
vacuum energy density $\rho_{\mathrm{vac}}\sim (2.4\times 10^{-3}\,\mathrm{eV%
})^4$ could explain the acceleration of the universe.

We can define an effective cosmological constant
\begin{equation}
\lambda_{\mathrm{eff}}=\lambda_0+\lambda_{\mathrm{vac}},
\end{equation}
where $\lambda_0$ is the ``bare'' cosmological constant in Einstein's
classical field equations, and $\lambda_{\mathrm{vac}}$ is the contribution
that arises from the vacuum density $\lambda_{\mathrm{vac}}=8\pi G\rho_{%
\mathrm{vac}}$. Already at the standard model electroweak scale $\sim 10^2$
GeV, a calculation of the vacuum density $\rho_{\mathrm{vac}}$, based on
local quantum field theory, results in a discrepancy of order $10^{55}$ with
the observational bound
\begin{equation}  \label{vacbound}
\rho_{\mathrm{vac}} \leq 10^{-47}\, (\mathrm{GeV})^4.
\end{equation}
This results in a severe fine-tuning problem of order $10^{55}$, since the
virtual quantum fluctuations giving rise to $\lambda_{\mathrm{vac}}$ must
cancel $\lambda_0$ to an unbelievable degree of accuracy. This is the
``particle physics'' source of the cosmological constant problem.

A model of nonlocal quantum gravity with the action $S=S_G+S_M$ was
presented in ref.~\cite{Moffat} with the gravitational action ($%
\kappa^2=32\pi G$)\footnote{%
A paper is in preparation in which more complete details of the model will
be provided.}:
\begin{equation}
S_G=-\frac{2}{\kappa^2}\int d^4x\sqrt{-g}\biggl\{R[g,\mathcal{G}%
^{-1}]+2\lambda_0\biggr\}.
\end{equation}
The matter action $S_M$ for the simple case of a scalar field $\phi$ is
given by
\begin{equation}
S_M=\frac{1}{2}\int d^4x\sqrt{-g}\mathcal{G}%
^{-1}\biggl(g^{\mu\nu}\nabla_\mu\phi\mathcal{F}^{-1}\nabla_\nu\phi -m^2\phi\mathcal{F}%
^{-1}\phi\biggr).
\end{equation}
Here, $\mathcal{G}$ and $\mathcal{F}$ are nonlocal regularizing, \textit{%
entire} functions and $\nabla_\mu$ is the covariant derivative with respect
to the metric $g_{\mu\nu}$. We chose the covariant functions
\begin{equation}
\mathcal{G}(x)=\exp\biggl[-\mathcal{D}(x)/\Lambda_G^2\biggr],
\end{equation}
\begin{equation}
\mathcal{F}(x)=\exp\biggl[-({\mathcal D}(x)+m^2)/\Lambda_M^2)%
\biggr],
\end{equation}
where $\mathcal{D}\equiv\nabla_\mu\nabla^\mu$, and $
\Lambda_G$ and $\Lambda_M$ are gravitational and matter energy scales,
respectively.

A calculation of the first-order vacuum polarization tensor $\Pi^{\mathrm{%
Gvac}}_{\mu\nu\rho\sigma}$ for the coupling of gravitons to a loop
consisting of photons yields for the vacuum energy density
\begin{equation}
\rho_{\mathrm{vac}}\sim M^2_{\mathrm{PL}}\Pi^{\mathrm{Gvac}}(p)\sim\Lambda_{%
\mathrm{Gvac}}^4.
\end{equation}
If we choose $\Lambda_{\mathrm{Gvac}}\leq 10^{-3}$ eV, then the quantum
correction to the bare cosmological constant $\lambda_0$ is suppressed
sufficiently to satisfy the observational bound on $\lambda$, \textit{and it
is protected from large unstable radiative corrections}.

This provides a solution to the cosmological constant problem at the energy
level of the standard model and possible higher energy extensions of the
standard model. The universal fixed gravitational scale $\Lambda_{\mathrm{%
Gvac}}$ corresponds to the fundamental length $\ell_{\mathrm{Gvac}}\leq 1$
mm at which virtual gravitational radiative corrections to \textit{pure}
vacuum energy are cut off. However, it is postulated in the model that
gravitons coupled to ordinary matter have a form factor that is controlled
by the energy scale, $\Lambda_{GM}=\Lambda_M\sim 1-10$ TeV, which avoids any
measurable violation of the WEP for gravitons coupled to ordinary matter and
guarantees that the calculations of standard model diagrams agree with
experiment. This would lead to a strong violation of the WEP for coupling of
low energy gravitons to pure zero-point vacuum energy compared to the
coupling to ordinary matter.

To explore the possibility that the cosmological constant arises from zero
point energy, Ross~\cite{Ross} has suggested a novel scenario in which the
violation of the WEP that would result from such an effect might be observed
in an E\"{o}tv\"{o}s experiment. In his proposal, the acceleration of a
spherical test mass of aluminum would be compared with that of a similar
test mass made from another material (e.g., a monel metal like copper or
silver). He chose aluminum because it has a relatively sharp transition from
reflectance to absorption of electromagnetic waves at photon energies of
approximately 15.5 eV. His analysis indicated that the magnitude of the
missing zero-point energy density inside aluminum is given by
\begin{equation}
\mathcal{E}=\frac{4\pi }{(\hbar c)^{3}}\int_{0}^{E_{\mathrm{max}}}E^{3}dE,
\end{equation}%
where $E_{\mathrm{max}}$ is the energy at which aluminum becomes
transparent. For $E_{\mathrm{max}}=15.5$ eV, one obtains $\mathcal{E}%
=2.37\times 10^{19}\,\mathrm{eV}/\mathrm{cm}^{3}$. The rest mass energy
density is $1.52\times 10^{33}$ eV, so he found that the ratio of the
zero-point energy density inside the aluminum sphere to the rest mass energy
density is $1.6\times 10^{-14}$. Therefore, if comparisons were made between
test masses of aluminum and, e.g., copper, a violation of WEP at
approximately this level should be observed if the zero point energy does
not couple to the gravitational field. Such an experiment would be a direct
test of the role that a purely quantum mechanical effect plays in general
relativity~\cite{Gillies}.

In a long series of elegant experiments with rotating torsion balances, the
\textit{E\"{o}t-Wash} Group has searched for composition dependence in the
gravitational force via tests of the universality of free fall. In terms of
the standard E\"{o}tv\"{o}s parameter $\eta $, they have reached
sensitivities of $\eta \sim 1.1\times 10^{-12}$ in comparisons of the
accelerations of Be and Al/Cu test masses~\cite{Su} and, more recently, have
resolved differential accelerations of approximately $1.0\times 10^{-14}\,%
\mathrm{cm/s^{2}}$ in experiments with other masses~\cite{Adelberger}.
Drop-tower experiments now underway in Germany~\cite{Koch} have as their
goal testing WEP at sensitivities of $\eta \sim 1\times 10^{-13}$, and
Unnikrishnan describes a methodology under study at the Tata Institute of
Fundamental Research in India wherein torsion balance experiments aiming at
sensitivities of $\eta \sim 1\times 10^{-14}$ are being developed~\cite%
{Unnikrishnan}

While it is not yet clear what ultimate sensitivities might be reached by
terrestrial experiments of these types, it is generally accepted that
significant gains in sensitivity will be made by space-based WEP
experiments, and several of them are presently in various stages of planning
and development. Among these are the STEP satellite~\cite{Everitt}, the
MICROSCOPE experiment~\cite{Touboul}, the Galileo Galelei (GG) mission ~\cite%
{Nobili}, and Project SEE~\cite{Sanders}. The target sensitivities of each
of them are STEP: $\eta \sim 10^{-18}$, MICROSCOPE: $\eta \sim 10^{-15}$,
GG: $\eta \sim 10^{-17}$, and SEE: $\eta \sim 10^{-16}$. The first three
missions are in advanced stages of planning and hardware testing, and are
expected to be launched over the next few years. Project SEE is still
undergoing rigorous conceptual evaluation and is not yet a scheduled
mission. However, one of the significant points of interest about it is the
relative simplicity of the test mass configuration: one large spherical mass
and one small one undergo a three-body interaction with the Earth during the
orbit of the capsule containing them~\cite{Sanders2}. Analysis of the small
mass motion then yields measures of WEP violation, the absolute value of the
Newtonian gravitational constant, and the time variation of it. Because it
is still an early stage endeavor, it might be possible to incorporate into
the mission schedule a test of WEP at the SEE target sensitivity using
aluminum and copper test masses, in order to examine the zero-point energy
coupling. This would avoid the need to design, fund and carry out a separate
independent mission, which would be very difficult to justify because of the
large costs. Further details about the SEE mission are given in the review
by Sanders and Gillies~\cite{Sanders3}.

The website http://www.phys.utk.edu/see/ provides a description of the
fundamental design features of the proposed SEE mission. It is meant to be a
multi-functional platform for gravitational physics experiments, providing
the capability to measure the strength of the gravitational interaction
between a pair of test bodies co-orbiting the Earth in a drag-free
environment. The nearly identical circular orbits of the test bodies in the
field of the Earth results in gravitationally-governed relative motions
between them of the type predicted first by George Darwin in 1897 and
exhibited in the celestial mechanics of two of the moons of Saturn [16].
When one of the masses is much smaller than the other, an analysis of the
mechanics of the interaction reveals how measurements of the relative
motions can lead to a determination of the absolute value of the Newtonian
gravitational constant and its time-rate of change, as well as limits on
violations of the weak equivalence principle. As with any high precision
measurement system of this type, great care must be taken to circumvent the
problems arising from competing effects, noise, and other sources of
experimental uncertainty. The satellite itself provides the first line of
defense, with its capsule consisting of a series of concentric cylindrical
shells that serve as radiation baffles and which provide the drag-free
containment for the test masses. The orbital configuration will be a
sun-synchronous trajectory that keeps the capsule illuminated at all times
and thus much closer to a state of constant thermal equilibrium. The large
and small test masses would be nominally 500 kg and 100 g respectively, and
small masses of different materials could be orbited simultaneously to make
differential measurements of the type needed to search for a WEP violation.
The details of the optimal orbital parameters, composition of the error
budget, and thermo-mechanics of the capsule are still under development.

If the SEE mission were able to accommodate a test of the type proposed
here, the level at which the WEP violation is predicted to occur, $\sim
1.6\times 10^{-14}$, means that a nearly 1\% resolution of it would be
possible within the context of the SEE interaction. In principle, this
degree of resolution would not only provide a definitive statement of the
presence of such an effect if it appeared at that level, but it would also
allow for enough sensitivity to observe weaker manifestations of variants of
the nominal prediction. The key issue in any case becomes one of maximizing
the signal size and, as always, minimizing the consequences of competing
effects. To accomplish the former task, a careful re-analysis of potential
test mass materials should be carried out to ensure that those with maximum
estimated difference in zero-point energy density are selected. This is an
important point, particularly in light of possible complexities that arise
in calculating the exact values of the Casimir effect cut-offs for spherical
mass configurations~\cite{Hagen}.

\qquad Another concern would be one of interpreting the meaning of an
unequivocal positive result. The task would be one of establishing, for
instance, whether or not the signal might alternatively be due to the
revelation of the presence of a long sought-for new weak force~\cite%
{Fischback}. Here again, selecting the proper experimental strategy within
the context of theoretical guidance would play the key role. Moreover, while
the prediction of the WEP violation in [5] is based on a straightforward
analysis of the vacuum energy density deficit in aluminum relative to that
present in other materials, a careful reconfirmation of that interpretation
will be important to establish that the level of the violation is indeed
what is claimed. \ A significant reduction of the strength of the expected
violation brought on by any reanalysis of the coupling of gravity to the
difference in vacuum energy densities in the test masses might place the
measurement below the reach of the SEE Mission's sensitivity.  Lastly, if a
satellite experiment were successful in revealing the predicted effect,
physics would be faced with one additional concern: the need to repeat and
independently confirm a finding of this significance. Perhaps completely new
technologies for terrestrial experiments could eventually be called on to
meet any such future need

\qquad As a point of historical interest, we note that one of the earliest
cases where vacuum polarization effects were discussed within the context of
a gravitational physics experiment was in Long's interpretation of his
non-null result for a breakdown in the inverse square law of Newtonian
gravity~\cite{Long}. He claimed to find a scale dependence in the Newtonian
gravitational constant, G and argued that a certain type of vacuum
polarization process could explain his positive results. He further argued
that the results consistent with no breakdown that were obtained in the null
experiments of others ~\cite{Spero} did not invalidate his findings of a
variation in G with inter-mass spacing, because of the lack of a polarizing
gravitational field in such experiments. Chen~\cite{Chen} extended the
argument and contended that vacuum polarization could even resolve the
differences between the null and the non-null results via a very weak
shielding mechanism that could arise from it, but the very stringent limits
on gravitational shielding phenomena~\cite{Unnikrishnan2,Unnikrishnan3}
derived from recent experiments block that possibility. The present
consensus is that the Newtonian inverse square law has been proven to be
valid over the ranges through which it has been tested experimentally. The
new laboratory searches for possible violations of it now focus on the
sub-millimetric regime, and are driven largely by predictions for
short-range variations in gravity that arise in extra-dimensional theories ~%
\cite{Dvali,Ritter}.

We call for a satellite test of the WEP that is designed to determine if the
cosmological constant arises from zero-point energy, and whether the WEP is
violated for coupling of gravitons to pure vacuum energy compared to
ordinary matter. An observed violation of the WEP would serve as
experimental support for the conjectured resolution of the cosmological
constant problem, based on a model of nonlocal quantum gravity and quantum
field theory~\cite{Moffat,Moffat2}.

\vskip0.2 true in \textbf{Acknowledgments} \vskip0.2 true in This work was
supported by the Natural Sciences and Engineering Research Council of
Canada. GTG was supported in part by a NASA subcontract from the University
of Tennessee to the University of Virginia, and he thanks D. K. Ross, A. J.
Sanders and C. S. Unnikrishnan for interesting discussions. \vskip0.5 true in

\end{document}